\documentclass[aps,prb,twocolumn,
               superscriptaddress,
               floatfix, 10pt]{revtex4-2}
\usepackage{amssymb}
\usepackage{amsmath}
\usepackage{braket}
\usepackage{tikz}
\usepackage{hyperref}
\usepackage{makecell}
\usepackage{caption}
\usepackage{bbm}

\hypersetup{colorlinks,
   linkcolor={blue!70!black},
   citecolor={blue!70!black},
   urlcolor= {blue!70!black}
}

\usetikzlibrary{calc}
\usetikzlibrary{arrows.meta}
\usetikzlibrary{decorations.markings}

\def\l{\left}
\def\r{\right}

\def\ola{{\overline{a}}}
\def\olb{{\overline{b}}}
\def\olc{{\overline{c}}}

\newcommand{\gr}[1]{{ \gamma^r_{#1}}}
\newcommand{\gb}[1]{{\gamma^b_{#1}}}
\newcommand{\sx}[1]{{ \sigma^x_{#1}}}
\newcommand{\sy}[1]{{\sigma^y_{#1}}}
\newcommand{\sz}[1]{{ \sigma^z_{#1}}}

\begin{document}

\title{Symmetry-Protected Topological Phases in the Triangular Majorana–Hubbard Ladder}

\author{Will Holdhusen}
\affiliation{Department of Physics and Astronomy, Western Washington University, Bellingham, WA 98225, USA}

\author{Alberto Nocera}
\affiliation{Stewart Blusson Quantum Matter Institute, University of British Columbia, Vancouver, BC V6T 1Z4, Canada}

\author{Jian-Xin Zhu}
\affiliation{Theoretical Division, Los Alamos National Laboratory, Los Alamos, New Mexico 87545, USA}

\author{Armin Rahmani}
\affiliation{Department of Physics and Astronomy, Western Washington University, Bellingham, WA 98225, USA}

\begin{abstract}
We uncover a family of symmetry-protected topological phases in a model of interacting Majorana fermions on triangular-lattice ladders. Included among these phases is one in which symmetry-protected topological order coexists with spontaneous symmetry breaking.
Our results indicate a mechanism through which interactions among Majorana modes generate nontrivial symmetry-fractionalized orders.
These phases may be realized in arrays of vortex-bound Majorana zero modes on the surface of topological superconductors.
\end{abstract}

\maketitle

\begin{figure}
\centering
\includegraphics[width=\linewidth]{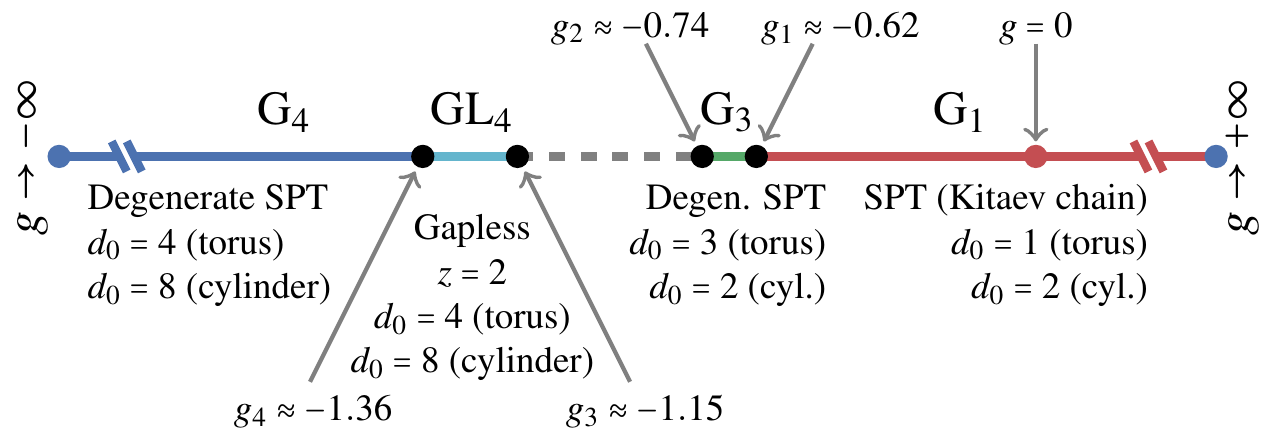}
\caption{
Phase diagram of the four-leg triangular-lattice Majorana-Hubbard ladder obtained from DMRG simulations. The four labeled phases are stable under finite-size scaling while the number and nature of the phases in the unlabeled region (illustrated with a dashed gray line) depends strongly on system size. Critical points $g_i$ were obtained from finite-size scaling on cylindrical geometries shown in Fig.~\ref{fig:obc_results}.
}
\label{fig:phase}
\end{figure}

\begin{figure}
\centering
\includegraphics[width=\linewidth]{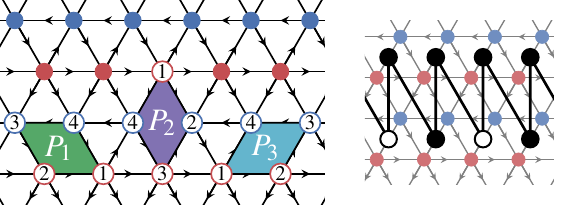}
\caption{
Decorated triangular lattice encoding hopping signs $\eta_{n,m}$ from Eq.~\eqref{eqn:h} as arrows. Red and blue circles indicate the two-site unit cell structure formed by this gauge. Numbered sites indicate the index order for plaquette operators. The 4-leg cylindrical ladder we study is formed by identifying the top and bottom rows of sites.
On the right, large circles indicate Dirac and spin-1/2 sites, with the bold path connecting them following the index $m$ defined in Eq.~\eqref{eqn:jw}. Filled and empty circles illustrate one of the ground states of $H_4$ as defined in Eq.~\eqref{eqn:h4}.
}
\label{fig:lattice}
\end{figure}

Symmetry-protected topological (SPT) phases form an important class of topological matter, and identifying microscopic models that support novel SPT phases remains a central challenge~\cite{Pollmann2010,Chen2011, Chen2011b, Senthil2015}.
Deep connections exist between fermionic SPT physics and Majorana zero modes (MZMs) \cite{Kitaev2001, Chen2011,Fidkowski2010,Fidkowski2011,Verresen2017}.
For this reason, models of interacting MZMs provide a framework for realizing a variety of SPT phases.
Here, we show that four-leg triangular-lattice ladders of interacting Majoranas host an unexpectedly rich phase diagram containing several distinct gapped phases with SPT order, including a novel degenerate SPT phase in which parity-protected Majorana edge modes coexist with spontaneous symmetry breaking (SSB) of microscopic symmetries. These findings enrich the landscape of phases that arise from strong interactions in Majorana systems and reveal new possibilities for symmetry-fractionalized quantum matter.

Here, we investigate SPT phases stabilized by the Majorana-Hubbard (MH) model on four-leg triangular-lattice ladders.
The MH model, with nearest-neighbor hopping and local four-site interactions, provides a minimal lattice description of interacting MZMs \cite{Rahmani2015, Rahmani2019b}, exhibiting rich phases in one and two dimensions \cite{Affleck2017, Rahmani2019a, Li2018, Kraus2011, Tummuru2021}. 
 Earlier studies of the triangular-lattice Majorana–Hubbard (TLMH) model reported a gapless phase in four-leg ladders, in contrast to mean-field predictions of a gapped phase \cite{Tummuru2021}. We show that the apparent gaplessness is a frustration-induced artifact of odd-length tori and uncover a much richer phase diagram, shown in Fig.~\ref{fig:phase}, that reveals novel SPT phases.

Specifically, we find the phase in the strongly attractive region, which we label G$_4$, is a gapped, degenerate SPT phase exhibiting both SSB and parity-protected Majorana edge modes. The SSB manifests as four distinct bulk ordering patterns related by exact microscopic symmetries of the Hamiltonian.
We support this result by adiabatically connecting the phase to a much simpler exactly solvable model that manifests its topological properties.

Likewise, we demonstrate that the G$_1$ phase, which extends from weakly attractive to strongly repulsive interactions, 
possesses fermionic SPT order and is adiabatically connected to the topological phase of the Kitaev chain \cite{Kitaev2001}.
Adjacent to the G$_1$ phase is one other gapped phase, G$_3$.
All three gapped phases exhibit signatures of fermion parity-protected topological physics including degenerate entanglement spectra \cite{Pollmann2010}, long-range string correlations \cite{Haldane1983,Affleck1987,Kennedy1992,Else2013,den1989,Pollmann2012,Sorensen2024}, and a switch in ground state parity between periodic and antiperiodic boundary conditions \cite{Kitaev2001, Ortiz2014, Lapa2020}.

Between the G$_3$ and G$_4$ phases, we observe
a gapless phase (GL$_4$) with dynamical critical exponent $z=2$. This exponent is characteristic of Lifshitz criticality \cite{Volovik2017}, which arises in systems including the one-dimensional spin-$1/2$ Heisenberg ferromagnet \cite{Bonner1987,Tarasevych2022}, gapless frustration-free Hamiltonians \cite{Masaoka2024}, and the Majorana-Hubbard chain, where the critical point instead has $z=3$ \cite{Rahmani2015}. Extended Lifshitz-critical regions are rare but have been found along critical lines \cite{Masaoka2024,Li2025b,Zhuang2026} and in field theories with a continuously tunable parameter \cite{Ardonne2004}.

Majorana fermions are thought to occur as zero-energy modes bound to vortex lattices on the surface of topological superconductors \cite{Ivanov2001, Biswas2013, Mishmash2019} or Fu--Kane interfaces in which a topological insulator is proximitized by an $s$-wave superconductor \cite{Fu2008}.
Experimental signatures consistent with vortex-bound MZMs have been found both in candidate $p$-wave materials \cite{wang2018evidence, kong2019half, Chiu2020} and in systems realizing the Fu--Kane platform \cite{xu2015experimental,liu2024signatures, Gaggioli2025}, although these results cannot yet rule out low-energy trivial excitations \cite{Ge2023, Machida2024}.
This latter case is especially promising as a potential realization of our model since tunable MZM hybridization allows for tunability of the ratio of interaction strength to hybridization amplitude and gives access to the strongly-interacting regime \cite{Chiu2015,Rahmani2015,Rahmani2015b}.

\section{Model}

The triangular-lattice Majorana-Hubbard (TLMH) model $H = H_0 + H_I$ introduced in \cite{Tummuru2021} consists of nearest-neighbor hopping and 4-fermion plaquette interactions:
\begin{equation}
H_0=it\sum_{\braket{p,q}}\eta_{p,q}\gamma_{p}\gamma_{q}, \quad
H_I=g\l(P_1 + P_2 + P_3\r),
\label{eqn:h}
\end{equation}
where $\braket{p,q}$ are nearest-neighbor sites on the triangular lattice.
The plaquette operators $P_\mu =\sum_{p_\mu} \gamma_1 \gamma_2 \gamma_3 \gamma_4$ are summed over each 4-site plaquette $p_\mu$ with orientation $\mu=\{1,2,3\}$ and index order shown in Fig.~\ref{fig:lattice}.
The antisymmetric matrix $\eta_{p,q}=\pm 1$ in \eqref{eqn:h} 
encodes the $\mathbb Z_2$ gauge freedom $\gamma \rightarrow -\gamma$ inherent to each Majorana mode.
The product of these phases around a closed loop defines a gauge-invariant quantity \cite{Grosfeld2006}. For Majorana zero modes bound to a vortex lattice, this quantity is equal to $-1$ for each triangular plaquette, corresponding to a $\pi$ flux \cite{Kraus2011,Liu2015}.  
One such configuration (see Refs.~\cite{Liu2015, Tummuru2021}) is shown in Fig.~\ref{fig:lattice}, with hopping along the arrows corresponding to $\eta=+1$. The lattice in Fig.~\ref{fig:lattice} has a two-site unit cell, shown as red ($\gamma^r$) and blue ($\gamma^b$) sites connected by the vector $\hat{\mathbf{c}}=(1,\sqrt{3})/2$. 
We label the unit cells by indices $(j,k)$, with positions $\mathbf{r}_{j,k}= j\hat{\mathbf{x}}+\sqrt{3}\,k\hat{\mathbf{y}}$ in units of the triangular lattice constant. The Hamiltonian in this basis is given explicitly in Appendix A.

This choice of unit cell leads to a natural description in terms of Dirac fermions
$c_{j,k} = \frac{1}{2}\l(\gamma_{j,k}^r + i \gamma_{j,k}^b\r)$. It is also useful to construct Pauli operators with the Jordan-Wigner (JW) transformation
\begin{equation}
\sigma^{x,y}_{m} =\pm\prod_{n=0}^{m-1}e^{i\pi c^\dagger_{n}c_n}\gamma^{r,b}_m,
\quad
m(j,k)=L_y j + k
\label{eqn:jw}
\end{equation}
with $\sigma_m^z = i\gamma_m^r \gamma_m^b = 2 c_m^\dagger c_m - 1$. Here and throughout, I am using zero-based indices $(j,k)$ where $j=0,\dots,L_x-1$ and $k=0,\dots,L_y-1$.

For the remainder of this paper, we focus on four-leg ladders with periodic boundary conditions along the shorter ($y$) direction ($y$-PBC), obtained by identifying the top and bottom rows of blue sites in Fig.~\ref{fig:lattice}, corresponding to the XC4 embedding defined in Ref.~\cite{Szasz2018}.
We consider both cylindrical geometries, with open boundaries along the $x$ direction ($x$-OBC), and toroidal geometries with periodic or antiperiodic boundary conditions along $x$ [$x$-(A)PBC]. The $4L_x$ Majorana modes in the ladder correspond to $N=2L_x$ spinless Dirac fermions. Unless otherwise noted, we restrict to even values of $L_x$ to avoid frustration induced by odd $L_x$ which produces the gapless mode identified in previous work \cite{Tummuru2021}. This frustration-induced gapless mode will be discussed later in this text.

The TLMH model preserves fermion parity, given by
\begin{equation}
    \mathcal P=(-1)^{\sum_{j,k} c_{j,k}^\dagger c_{j,k}}=\prod_m \sigma_m^z
    \label{eqn:parity}
\end{equation} 
The model inherits the rotational and translational symmetries of the triangular lattice, which become manifest when accompanied by appropriate $\mathbb{Z}_2$ gauge transformations. It also exhibits two antiunitary symmetries $\mathcal A_{x,y} = \Theta \mathcal R_{x,y}$ where $\mathcal R_{x,y}$ generates reflections about the $x$ and $y$ Cartesian axes and $\Theta$ is complex conjugation in the Majorana basis taking $i\rightarrow -i$ and $\gamma \rightarrow \gamma$ \cite{Tummuru2021}. 

Of particular importance for us are three symmetries that remain exact for $x$-OBC on the 4-leg ladder. These are the antiunitary symmetry $\mathcal A_x$, the unitary transverse translation $\mathcal T_y$, and a second antiunitary symmetry $\mathcal S$ combining longitudinal reflection with translation and a gauge transformation. On the ladder, these act on Majoranas as
\begin{subequations}
\label{eqn:symm_action}
\begin{align}
\mathcal T_y \gamma_{j,k}^{r,b}\mathcal T_{y}^{-1} =& -\gamma_{j,1-k}^{r,b}\\
\mathcal A_x \gamma_{j,k}^r \mathcal A_x^{-1} =& (-1)^j \gamma_{j,k}^r \\
\mathcal A_x \gamma_{j,k}^b \mathcal A_x^{-1} =& (-1)^{j+1} \gamma_{j,1-k}^b \\
\mathcal S \gamma_{j,k}^r \mathcal S^{-1} =& (-1)^{j}\gamma^b_{L_x-1-j, k}  \\
\mathcal S \gamma_{j,k}^b \mathcal S^{-1} =& (-1)^{j}\gamma^r_{L_x-1-j, 1-k}. 
\end{align}
\end{subequations}
For $L_x$ even and with an appropriate choice of overall phases, these obey
\begin{align}
\mathcal A_x^2 = \mathcal T_y^2  = 1,
\quad \mathcal S^2 = T_y, \quad \mathcal A_x \mathcal S \mathcal A_x^{-1}=\mathcal P \mathcal S^{-1}.
\end{align}
Fermion parity $\mathcal P$ commutes with all three symmetry operators, and the resulting group is a nontrivial extension of the dihedral group $D_4$ by fermion parity. These relations and further discussion of the symmetry group are present in Appendix B. As we will discuss, $\mathcal S$ is spontaneously broken in the G$_4$ phase.

\begin{figure}
    \centering
    \includegraphics[width=0.95\linewidth]{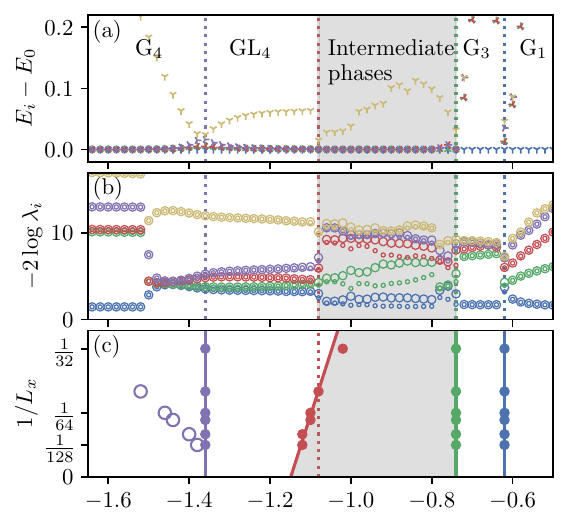}
    \caption{
    (a) Low-lying gaps $E_i-E_0$ calculated in the even parity sector of an $L_x=48$ cylinder.
    (b) ground state entanglement spectrum from the same system. Values are marked with alternating small and large circles to make degeneracies visible in the G$_1$, G$_3$, GL$_4$, and G$_4$ phases.
    (c) Finite-size extrapolation of critical couplings based on the location of many-body gap closures. Open circles indicate the location of the discontinuities in the entanglement spectrum near the GL$_4$-G$_4$ transition.}
    \label{fig:obc_results}
\end{figure}

Results for finite-size systems were obtained from DMRG based on the ITensor package \cite{ITensor, ITensorRelease}. With bond dimensions up to 2500, truncation errors were kept below $10^{-8}$ in gapped phases and increased to $\sim 10^{-5}$ near the critical points indicated in Figs.~\ref{fig:phase} and \ref{fig:obc_results}. To access the thermodynamic limit directly, we used variational uniform matrix product states (VUMPS) \cite{Vanderstraeten2019} to obtain infinite-MPS ground states for the G$_1$ and G$_4$ phases. Exact-diagonalization techniques were used for validation on small systems.

\begin{figure}
    \centering
    \includegraphics[width=.95\linewidth]{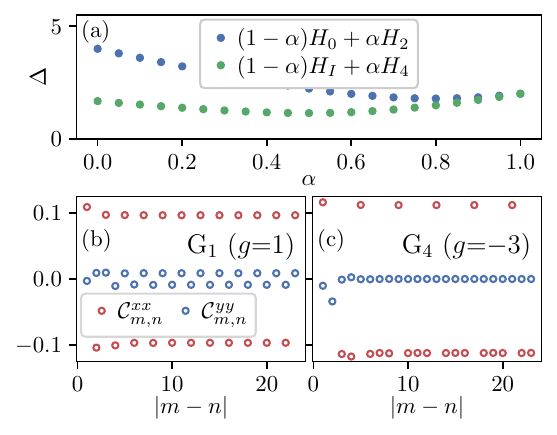}
    \caption{(a) Gap (extrapolated to the thermodynamic limit from finite-size $x$-OBC results) along a path connecting the noninteracting Hamiltonian $H_0$ to the Ising model $H_2$ and the interaction Hamiltonian $H_I$ to $H_4$.
    (b) String-order correlations $\mathcal C^{xx}_{m,n}$ and $\mathcal C^{yy}_{m,n}$ in the even-parity ground state of the G$_1$ phase on an $L_x=96$ cylinder. 
    Correlations are calculated with $m$ taken as the midpoint of the system. 
    (c) Correlations from the same system in 
    the G$_4$ phase.}
    \label{fig:g1g4}
\end{figure}

\section{Phase diagram}

\subsection{Noninteracting point and the G$_1$ phase}

The G$_1$ phase includes the noninteracting point $g=0$ with Hamiltonian $H=H_0$ [see Eq.~\eqref{eqn:h}] and can be readily diagonalized via a Bogoliubov transformation \cite{Tummuru2021}. Additionally, we find that the TLMH at this point may be adiabatically deformed into a fermionic Hamiltonian $H_2$ equivalent to the 1-d Ising model under JW transformation:
\begin{align}
H_2 =&-it\sum_j\l(\gamma_{j,0}^b\gamma_{j,1}^r + \gamma^b_{j,1}\gamma^r_{j+1,0}\r) \nonumber\\
    =&t\sum_{m}\sigma_m^x \sigma_{m+1}^x.
\end{align}
Here, $m=m(i,j)$ is the index defined in \eqref{eqn:jw}. This connection is demonstrated in Fig.~\ref{fig:g1g4}, in which the gap, plotted as a function of the parameter $\alpha$, remains finite as we interpolate between $H_0$ and $H_2$.

With open boundaries, $H_2$ is known to host a Majorana edge mode at each end of the chain, reflected in the absence of $\gamma_{0,0}^r$ and $\gamma_{L_x-1,1}^b$ from the Hamiltonian \cite{Kitaev2001}.
These MZMs characterize the phase and are robust as long as the gap remains finite. In Appendix C, we show that this phase terminates at a first-order transition in the limit $g \to +\infty$, approached by taking $t \to 0$ at fixed $g>0$. 
We also observe the fermion parity switch characteristic of the Kitaev phase \cite{Kitaev2001} 
in which ground state parity $\braket{\mathcal P}$ changes from $+1$ to $-1$ as the boundary conditions are taken from $x$-APBC to $x$-PBC. The resulting product  $\braket{\mathcal P_{\rm APBC}}
\braket{\mathcal P_{\rm PBC}}=-1$ is consistent with a nontrivial
fermionic $\mathbb Z_2$ parity-switch invariant \cite{Ortiz2014, Lapa2020}.
This phase also exhibits the degenerate entanglement spectrum characteristic of SPT phases \cite{Pollmann2010}, as shown in Fig.~\ref{fig:obc_results}.

In spin language, the ground state of $H_2$ exhibits long-range correlations in $\sigma^x$. To see these in the TLMH, we apply the JW transformation \eqref{eqn:jw} to write them in terms of fermion strings: $C^{xx}_{m,n}=(-1)^{n-m}\braket{\gamma_{m}^r \sigma_{m}^z\sigma_{m+1}^z \dots \sigma_{n-1}^z \gamma_{n}^r}$.
These correlations are plotted in Fig.~\ref{fig:g1g4}. We also observe weaker correlations in $\sigma^y$ (obtained by substituting $\gamma^b$ for $\gamma^r$), consistent with other terms appearing in the JW transformation of $H_0$ shown in Appendix A.
Connected correlations in local fermionic operators decay exponentially in this phase.

\subsection{G$_4$ phase and strong-coupling limits}
As indicated by Fig.~\ref{fig:phase}, we find that the G$_4$ phase remains stable as $g\rightarrow -\infty$. Results, obtained by taking $t\rightarrow 0$ with $g<0$, are demonstrated in Appendix C. Due to a gauge transformation connecting the TLMH in both infinite-coupling limits \cite{Tummuru2021}, we thus also identify the $g\rightarrow +\infty$ limit (at which point the G$_1$ phase terminates in a first-order transition) with the G$_4$ phase.

In either limit, it is again possible to adiabatically connect the TLMH to a simple Hamiltonian (see Fig.~\ref{fig:g1g4}), this time with four-spin interactions:
\begin{align}
H_4
=&g\sum_j\l(
\gamma^b_{j,0}\gamma^r_{j,1}\gamma^b_{j+1,0}\gamma^r_{j+1,1}
+
\gamma^b_{j,1}\gamma^r_{j+1,0}\gamma^b_{j+1,1}\gamma^r_{j+2,0}\r)
\nonumber\\
 =&-g\sum_{m}\sigma^x_{m}\sigma^x_{m+1}\sigma^x_{m+2}\sigma^x_{m+3}.
\label{eqn:h4}
\end{align}
The four-spin Ising-like interactions of $H_4$ have been shown to be equivalent to the 8-state Potts model \cite{Blote1986, Alcaraz1986}. Also see Refs.~\cite{Bonfim2006, Turban2016} for discussion of spin chains combining $H_2$ and $H_4$ with a transverse field. 
Interestingly, the one-dimensional Majorana-Hubbard chain also hosts phases connected to $H_4$ \cite{Rahmani2015}. 

The stability (instability) of the $G_4$ phase with $g<0$ ($g>0$) in the presence of hopping terms proportional to $H_0$ can be understood from the classical eigenstates of $H_2$ and $H_4$. 
For $g>0$, the four-spin term in $H_4$ has negative coefficient, and the eightfold-degenerate ground state manifold consists of repeated product states of local $\sigma^x$ eigenstates $\ket{\pm}$ (e.g., $\ket{++++}$, $\ket{+-+-}$, $\ket{++--}$, $\ket{+--+}$) and their partners obtained by applying $\mathcal P=\prod_m\sigma^z_m$.
Notably, two of these states are also ground states of $H_2$, so any infinitesimal $H_2$ lifts the degeneracy and destabilizes the $G_4$ phase. 

On the other hand, for $g<0$, the ground states can be formed as tensor products of four-spin configurations $\ket{+---}$ and permutations thereof, as illustrated in Fig.~\ref{fig:lattice}. (In the full TLMH, $\sigma^x$ correlations showing these patterns appear in the $G_4$ phase, as shown in Fig.~\ref{fig:g1g4}.)
Unlike the $g>0$ case, small $H_2$ does not lift the degeneracy. This is consistent with numerical studies of the $H_2$–$H_4$ model \cite{Bonfim2006, Turban2016}. 

As is the case in $H_2$, the first and last Majorana along the one-dimensional path are absent from $H_4$ and thus correspond to zero-energy edge modes protected by fermion parity $\mathcal P$. As in the G$_1$ phase, we find further indications of SPT order including a  fermion parity switch and degenerate entanglement spectrum throughout the phase (as shown in Fig.~\ref{fig:obc_results}), which in the presence of an SSB-induced degenerate ground state is consistent with either symmetry fractionalization \cite{Pollmann2010} or else a fine-tuned cat-state superposition. In our case, we are able to rule out the latter case as the infinite MPS transfer matrix (obtained via VUMPS) has a unique dominant eigenvalue \cite{Garcia2008, Pollmann2012}.

\begin{figure}
    \centering
    \includegraphics[width=.95\linewidth]{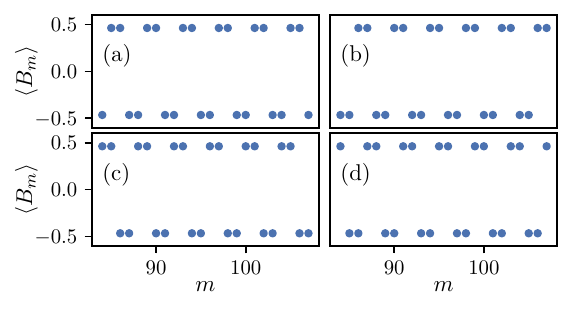}
    \caption{Expectation values of the bilinears $B_m$ defined in Eq.~\eqref{eqn:bilinears} computed in the G$_4$ phase, specifically at $g=-3$ in an $L_x=96$ cylinder. Subplots (a)-(d) correspond to the four even-parity states in the ground state manifold, which are distinguishable by local expectation values $\braket{B_m}$.}
    \label{fig:majorana_bilinears}
\end{figure}

The presence of these parity-protected MZMs accounts for the observed degeneracy between the even and odd parity sectors in the G$_4$ phase with $x$-OBC. The remaining fourfold ground state degeneracy within each sector is due to SSB, which can be seen directly in local fermionic observables.
In particular, Fig.~\ref{fig:majorana_bilinears} shows that the four symmetry-broken configurations are resolved by the Majorana bilinears
\begin{equation}
    B_{m}=i{\gamma_{m}^b\gamma_{m+1}^r} \label{eqn:bilinears}
\end{equation}
using the indexing established in Eq.~\eqref{eqn:jw}.
As discussed in Appendix D,
exact diagonalization confirms that the microscopic symmetry $\mathcal S$ introduced in Sec.~I cyclically permutes all four symmetry-broken states within a fixed parity sector.

\subsection{Additional phases}
With our understanding of the low and strong-coupling limits, we now discuss the remaining phases we identify in the TLMH.
Our discussion here focuses on simulation results from cylinders with $L_x$ ranging from 32 to 128, with key results plotted in Fig.~\ref{fig:obc_results}. Qualitatively similar results obtained on even-$L_x$ tori are discussed in Appendix E.

The fermionic SPT G$_1$ phase extends from a first-order transition at $g\rightarrow \infty$ to a phase transition at $g_1\approx -0.62$. 
Below this point, we observe the G$_3$ phase. Like G$_1$, this phase exhibits a fermion parity switch between $x$-APBC and $x$-PBC. With $x$-APBC, the three ground states have
even parity, whereas with $x$-PBC, they have odd parity. In both
cases, a finite gap separates the ground state manifold from the
lowest state of opposite parity. With $x$-OBC, there is a unique
ground state in each parity sector, and these states are degenerate,
yielding a twofold ground state degeneracy.

The G$_3$ phase also exhibits degeneracies in the entanglement spectrum (seen in Fig.~\ref{fig:obc_results}) and long-range string-order correlations discussed in Appendix F. Together, these results are consistent with a fermion-parity-protected topological component. They do not, however, provide a complete symmetry classification of the phase and cannot account for the threefold degeneracy present on the torus.

Another transition at $g_2\approx -0.74$ brings us out of the G$_3$ phase into a region characterized by strong entanglement, which makes obtaining high-quality DMRG results difficult. Unlike the phases enumerated in Figs.~\ref{fig:phase} and \ref{fig:obc_results}, the phases occurring in this region do not occur consistently across different system sizes and upon changing boundary conditions. 
Such inconsistencies may hint at an ordered phase with a large unit cell that leads to incommensurability effects.
In any case, we leave detailed examination of this region for future work.

Decreasing $g$ further brings us to a phase transition at $g_3$ leading to the GL$_4$ phase, which is easily distinguished by its fourfold degeneracy on the torus (doubled on the cylinder), degenerate entanglement spectrum, and on the torus, a change in ground state parity relative to the adjacent phases. This phase also exhibits a change in ground state fermion parity between $x$-PBC and $x$-APBC. Correlation functions in this phase are plotted in Appendix F.

Unlike the other named phases, the GL$_4$ phase appears to be gapless:
Finite-size scaling presented in Appendix E indicates the bulk gap $\Delta$ above the ground state manifold behaves as $\Delta\propto 1/L_x^2$ in the thermodynamic limit, consistent with dynamical critical exponent $z=2$. At the same time, the phase manifests the same degenerate entanglement and entanglement entropy characteristic of SPT physics, indicating possible gapless SPT physics \cite{Li2024}.
DMRG results on cylinders indicate a phase transition from the GL$_4$ phase into the previously discussed G$_4$ phase at $g_4\approx -1.36$, signaled by a  many-body gap closure (seen in Fig.~\ref{fig:obc_results}) and a discontinuity in the third derivative of energy density discussed in Appendix E.

An interesting finite-size anomaly occurs at the GL$_4$-G$_4$ transition: while the gap closure occurs at $g\approx -1.36$ for all system sizes considered, discontinuities in the ground state entanglement spectrum, which generally accompany phase transitions \cite{Sorensen2024}, occur generally at more negative couplings ($g\approx -1.5$ for the $L_x=48$ results in Fig.~\ref{fig:obc_results}).
 This would seem to be due to the fact that the gap closing doesn't, in a finite system, involve the true ground state due to finite-size splitting, but instead has an immediate effect only on higher-energy states within the ground state manifold. As indicated by open circles in subfigure (c) of Fig.~\ref{fig:obc_results}, the coupling at which these discontinuities occur approaches $-1.36$ as $L_x$ increases.

\subsection{Frustration-induced gapless mode}
Throughout the preceding analysis we assumed $L_x$ to be even. On a cylinder ($x$-OBC), no appreciable difference is observed between even and odd $L_x$. However, on odd-$L_x$ tori (as studied in Ref.~\cite{Tummuru2021}), the G$_4$ phase is replaced by a critical phase with relativistic dispersion ($\Delta\propto 1/L$) \cite{Tummuru2021} above a twofold degenerate ground state.
Insight from the preceding discussion of the G$_1$ and G$_4$ phases helps account for this discrepancy.

The G$_1$ phase occurs on both even and odd $L_x$ tori over essentially the same range of couplings $g$. This is unsurprising, since the magnetic order in this phase is commensurate for both even and odd $L_x$ (the total number of spins $N=2L_x$ remains even in either case). More precisely, this commensurability holds when the spin Hamiltonian has open or periodic boundary conditions, whereas antiperiodic boundary conditions force the formation of a domain wall \cite{Dong2016}. The APBC case in the G$_1$ phase can be discarded by considering the  JW mapping introduced in Eq.~\eqref{eqn:jw}. Under this transformation, the even-parity ground state of the $x$-APBC fermionic Hamiltonian and the odd-parity ground state of the $x$-PBC fermionic Hamiltonian both correspond to periodic boundary conditions in the spin Hamiltonian.

The G$_4$ phase has a four-spin unit cell (inherited from the previously-discussed ground states of $H_4$ previously) that is incommensurate when the number of spins, $2 L_x$, is not a multiple of four.
Again, this incommensurability forces the formation of a domain wall. In the classical spin Hamiltonian $H_4$ (which, as in the previous case, we can always take to have either OBC or PBC in the ground state sector), this induces a degeneracy $d_0\propto L_x$. The additional interactions present in the full TLMH then delocalize the domain wall through hybridization, leading to a gapless excitation spectrum consistent with results obtained through finite-size scaling of DMRG results \cite{Tummuru2021}.
Similar frustration-induced gapless modes occur in other triangular lattice models when placed on toric and cylindrical geometries \cite{Green2000, Green2002}
as well as in one-dimensional spin models with periodic boundary conditions \cite{Cador2005, Campostrini2015, Dong2016, Florek2016, Giampaolo2019, Maric2020}.

Consistent with our identification of the gapless mode with a single defect, we show in Appendix G that odd-$L_x$ tori exhibit an $\mathcal O(1/L_x)$ increase in energy density compared to similarly-sized even-$L_x$ tori. This is consistent with a constant $\mathcal O(1)$ energy increase induced by the defect.

Unlike what should be seen from an isolated defect, the ground state in this region exhibits entanglement entropy scaling consistent with a conformal central charge $c=1$ \cite{Tummuru2021}, contrasting with the non-critical scaling associated with a single deconfined defect.
Following the same fitting procedure \cite{Calabrese2009}, we extend these results from $L_x=25$ to $L_x=65$ in Appendix G. 

As in the case of the GL$_4$ phase, we observe both degenerate entanglement spectrum and a fermion parity switch in this region. Again, these could indicate gapless SPT physics \cite{Li2024}.

\section{Conclusions}
We have performed a comprehensive numerical study of the TLMH on four-leg ladders under various boundary conditions, using large-scale DMRG and VUMPS in the computationally challenging attractive regime. We show that the previously reported gapless phase \cite{Tummuru2021} arises from specific choices of boundary conditions and system size. Our results reveal a rich phase diagram, featuring a sequence of gapped phases, including a conventional fermionic SPT and a phase that combines SPT order with spontaneous symmetry breaking. This work underscores the crucial role of boundaries in stabilizing exotic phases of interacting Majorana fermions in quasi-1D triangular-lattice geometries.
\begin{acknowledgments}
We thank Claudio Chamon, Lucasz Fidkowski, and Chris Laumann for helpful discussions. This work was primarily supported by DOE Grant No. DE-SC0024641 (W.H., A.R., and J.-X.Z.), which supported the conceptual development of the project, formulation of the theoretical framework, development of the model and computational approach, and analysis and interpretation of the results. Computations were in part supported by Center for Integrated Nanotechnologies, a DOE Basic Energy Sciences user facility, in partnership with the LANL Institutional Computing Program for computational resources. Work at Los Alamos National Laboratory (LANL) was carried out under the auspices of the U.S. Department of Energy (DOE) National Nuclear Security Administration (NNSA) under Contract No. 89233218CNA000001. 
\end{acknowledgments}

\appendix

\section{Details of the Triangular-lattice Majorana-Hubbard model}
We consider the triangular-lattice Majorana-Hubbard model (TLMH) given by Eq.~(1) with the phases $\eta_{p,q}=\pm 1$ fixed as depicted in Fig.~2. 

Expressed in the Majorana basis, the noninteracting portion of the TLMH is
\begin{align}
H_0 = it \sum_{j,k} \bigg[&
\gamma_{j,k}^r \l(\gamma_{j+1,k}^r + \gamma_{j,k}^b - \gamma_{j-1,k}^b\r)
 \\
&- \gamma_{j,k}^b\l(\gamma_{j+1,k}^b + \gamma_{j+1,k+1}^r + \gamma_{j,k+1}^r\r)
\bigg].\nonumber
\label{eqn:h0}
\end{align}
and the interactions are given by
\begin{subequations}
\begin{align}
P_1 = \sum_{j,k} 
\big(&\gamma_{j,k}^r \gamma_{j-1,k}^r \gamma_{j-2,k}^b \gamma_{j-1,k}^b \nonumber\\
& + \gamma_{j,k}^b \gamma_{j-1,k}^b \gamma_{j-1,k+1}^r \gamma_{j,k+1}^r\big)\\
P_2 = \sum_{j,k} 
\big(&\gamma_{j,k}^r \gamma_{j-1,k-1}^b \gamma_{j,k-1}^r \gamma_{j,k-1}^b \nonumber \\
& + \gamma_{j,k}^b \gamma_{j,k}^r \gamma_{j,k-1}^b \gamma_{j+1,k}^r\big)\\
P_3 =\sum_{j,k}
\big(&\gamma_{j,k}^r \gamma_{j+1,k}^r \gamma_{j+1,k}^b \gamma_{j,k}^b \nonumber\\
& + \gamma_{j,k}^b \gamma_{j+1,k}^b \gamma_{j+2,k+1}^r \gamma_{j+1,k+1}^r\big)
\end{align}
\end{subequations}
with total interaction $H_I = g \l(P_1 + P_2 + P_3\r)$.
Indices $(j,k)$ correspond to the unit cells depicted in Fig.~2 of the main text, which form a rectangular Bravais lattice.
Mapping the Hamiltonian to Dirac fermions using $c_{j,k}=\frac{1}{2}\l(\gamma_{j,k}^r +i\gamma^b_{j,k}\r)$ is straightforward.

\subsection{Jordan-Wigner transformation}

We can apply the Jordan-Wigner (JW) transformation defined in Eq.~\eqref{eqn:jw} to our Hamiltonian to form a corresponding spin-1/2 Hamiltonian. 
With $L_y=2$ for our 4-leg ladder, the noninteracting part of the Hamiltonian translates to
\begin{align}
H_0
=&
t\sum_m\big(
\sz{m} +\sx{m}\sx{m+1} 
+\sy{m}\sz{m+1}\sx{m+2}
\nonumber\\
& \qquad
+\sx{m}\sz{m+1}\sy{m+2}
+\sx{m-2}\sz{m-1}\sx{m}
\big) \nonumber\\
&+t\sum_{m\text{ even}}
\big(\sx{m}\sz{m+1}\sz{m+2}\sx{m+3}
+
\sy{m}\sy{m+1}\big).
\end{align}
The interaction terms become
\begin{subequations}
\begin{align}
P_1 
=
&
-\!\!\!\sum_{m \text{ even}}\!\!\!\l(
\sx{m}
\sx{m+1}
\sx{m+2}
\sx{m+3}
+
\sy{m}
\sy{m+1}
\sy{m+2}
\sy{m+3}
\r)\nonumber\\
&-\sum_m \sx{m}\sz{m+1}\sz{m+3}\sx{m+4} \\
P_2
=&
-\!\!\!\sum_{m \text{ even}}\!\!\!\l(
\sz{m}\sx{m+1}\sx{m+2}
+\sx{m}\sz{m+1}\sx{m+3}
+\sx{m}\sz{m+2}\sx{m+3}
\r)\nonumber \\
&-\!\sum_{m \text{ odd}}\!\sx{m}\sx{m+1}\sz{m+2} \\
P_3
=&-\!\!\!\sum_{m \text{ even}}\!\!\!\sx{m}\sz{m+1}\sy{m+2}
\sy{m+3}\sz{m+4}\sx{m+5} \nonumber \\
& -\sum_m \sz{m}\sz{m+2}
-\!\sum_{m \text{ odd}}\!
\sx{m}\sx{m+1}\sx{m+2}\sx{m+3}.
\end{align}
\end{subequations}

\section{Symmetry analysis}

Since every term in the TLMH is the product of an even number of Majorana fermions, it conserves total fermion parity $\mathcal P$ defined in Eq.~\eqref{eqn:parity}.
As established in Ref.~\cite{Tummuru2021}, the TLMH on an infinite lattice is symmetric under triangular-lattice translations $\mathcal T_a$, $\mathcal T_b$, and $\mathcal T_c$ along the directions $\hat a = \hat x$, $\hat b = \frac{1}{2}\l(-\hat x + \sqrt{3}\hat y\r)$, and $\hat c = \frac{1}{2}\l(\hat x + \sqrt{3}\hat y\r)$ so long as the translation is accompanied by an appropriate gauge transformation. Since these are not linearly independent operations, we consider just $\mathcal T_a$ and $\mathcal T_c$. 
The model also preserves antiunitary reflection symmetries $\mathcal A_x = \Theta \mathcal R_x$ and $\mathcal A_y = \Theta \mathcal R_y$, where $\Theta$ is the complex conjugation operator in the Majorana basis taking $i\rightarrow -i$ and $\gamma\rightarrow \gamma$ and $\mathcal R_{x,y}$ generate reflections about the $x$ and $y$ axes, which we take to pass through sites $\gamma_{i,0}^r$ and $\gamma_{0,j}^r$ respectively. 
These symmetries, and their actions on Majorana operators, are given in Ref.~\cite{Tummuru2021}.

Not all of these symmetries hold on the 4-leg cylindrical geometry we focus on. Since the observed degeneracies of the $G_4$ phase within a fixed parity sector occur with both $x$-PBC and $x$-OBC, we are primarily interested in just those symmetries preserved on the 4-leg cylinder.
To make this discussion precise for finite 4-leg cylinders, we need to fix indexing conventions. For the following, unit cells have indices $(i,j)$ where $i = 0,\dots, L_x-1$ and $j=0,1$.
It can be seen that on the 4-leg cylinder, of the preceding symmetries only $\mathcal A_x$ persists, while combinations of the others also generate symmetries of the model on the cylinder.

One such combination is the antiunitary operator $\mathcal S$, which can be written
\begin{equation}
    \mathcal S = T_a^{L_x-1}T_c \mathcal R_y \Theta. \label{eqn:S}
\end{equation}
This operator can be understood as performing the antiunitary reflection $\mathcal A_y$ about the left edge of the cylinder, followed by translation in the $\hat c$ direction to restore the original boundary termination and in the $\hat a$ direction to re-center the system.
Translation in the $y$ direction, generated by 
$\mathcal T_y = \mathcal T_c^2 \mathcal T_a^{-1}$, is another such operator with a simpler physical description. As we demonstrate shortly, $\mathcal T_y$ and $\mathcal S$ are not independent, and so we focus on the latter. 
The action of these operators on local Majoanas is given by Eq.~\eqref{eqn:symm_action} in the main text.

It can be seen that $\mathcal A_x$ and $\mathcal S$ are exact symmetries of the TLMH on the cylinder. As an example, consider their effects on $i\gamma_{j,k}^r\gamma_{j+1,k}^r$, the first term appearing in $H_0$ as written in Eq.~\eqref{eqn:h0}:
\begin{align*}
 \mathcal A_x \left(i\gamma_{j,k}^r\gamma_{j+1,k}^r\right)\mathcal A_x^{-1}
 =&i \gamma_{j,k}\gamma_{j+1,k}\\
  \mathcal S \left(i\gamma_{j,k}^r\gamma_{j+1,k}^r\right)\mathcal S^{-1}
 =&
-i\gamma_{L_x-1-j,k}^b \gamma_{L_x-2-j,k}^b.
\end{align*}
These are terms also appearing in $H_0$. Repeating this calculation for each term shows all three operators are symmetries of $H_0$. Similar calculations show the operators also preserve $H_I$.

\subsection{Group structure}
The operators $\mathcal A_x$, $\mathcal S$, and parity $\mathcal P$ form a group that is a nontrivial central extension of the dihedral group $D_4$ by fermion parity in the case of even $L_x$, with a related structure in the odd $L_x$ case.
As is shown in the following text, when $L_x$ is even, the operators satisfy the algebra
\begin{align}
\mathcal A_x^2 = 1, \quad 
\mathcal S^4 =1, \quad \text{and}\quad
\mathcal A_x \mathcal S \mathcal A_x^{-1} = \mathcal P \mathcal S^{-1}
\end{align}
with $[\mathcal P, \mathcal A_x] = [\mathcal P, \mathcal S] = 0$ and $\mathcal P^2=1$. 

To derive these relations, it is convenient to work with fermionic operators
corresponding to symmetric and antisymmetric orbitals on the rungs of the ladder:
\begin{equation}
d_{i} = \frac{1}{\sqrt 2}(c_{i,0}+c_{i,1})
\quad
\text{and}
\quad
a_{i} = \frac{1}{\sqrt 2}(c_{i,0}-c_{i,1}).
\end{equation}
These satisfy the Dirac fermion anticommutation relations, including $\{d_i, a_j\}=0$.
Under the symmetry operators, the new fermions transform as 
$\mathcal A_x d_j \mathcal A_x^{-1} =(-1)^j d_j$,
$\mathcal A_x a_j \mathcal A_x^{-1} = (-1)^j a_j^\dagger$,
$\mathcal S d_j \mathcal S^{-1} = -i (-1)^j d_{L_x-1-j}$,
and
$\mathcal S a_j \mathcal S^{-1} = i (-1)^j a^\dagger_{L_x-1-j}$.

Since $\mathcal A_x^2\gamma_{i,j}\mathcal A_x^{-2} = \gamma_{i,j}^{r,b}$, we see that $\mathcal A_x^2 = e^{i\alpha}$. The antiunitarity of $\mathcal A_x$ fixes $e^{i\alpha}=\pm 1$, but the phase cannot be absorbed by redefinition and instead 
must be fixed. This can be accomplished with reference to the Dirac vacuum state $\ket 0$ where $c_{i,j}\ket{0}=0$. 
Define $\ket{\Psi} \equiv \mathcal A_x \ket{0}$. 
Since $d_j\ket{0}=0$,
we must have $\mathcal A_x d_j \mathcal A_x^{-1} \mathcal A_x\ket{0}=0$. Using the definition of $\ket{\Psi}$ and the transformation of $d_j$, that means $d_j\ket{\Psi}=0$. Similarly, we get $a_j^\dagger \ket{\Psi} = 0$. This is satisfied for
\begin{equation}
\mathcal A_x \ket{0} = e^{i\theta_x}\prod_{j=0}^{L_x-1}a_j^\dagger \ket{0} \equiv e^{i\theta_x} \ket{\Phi}. \label{eqn:phi}
\end{equation}
where $\theta_x$ is a free parameter we may set as $\theta_x=0$.
Applying $\mathcal A_x$ again yields
\begin{equation*}
\mathcal A_x \ket{\Phi}
=\prod_{j=0}^{L_x-1}  \left(\mathcal A_x  a_j^\dagger \mathcal A_x^{-1} \right) \mathcal A_x\ket{0} 
=\prod_{j=0}^{L_x-1}  \left(\mathcal A_x  a_j^\dagger \mathcal A_x^{-1} \right)
\ket{\Phi}. 
\end{equation*}
Note how $a_j^\dagger$ transforms: $\mathcal A_x a_j^\dagger \mathcal A_x^{-1} = (-1)^j a_j$.
So,
\begin{equation*}
\mathcal A_x \ket{\Phi} =
\prod_{j=0}^{L_x-1}(-1)^j a_j \prod_{k=0}^{L_x-1} a_k^\dagger \ket 0 = \ket{0}.
\end{equation*}
In the second equality, the sign acquired from fermion anticommutation is cancelled by the sign of $\prod_{j=0}^{L_x-1} (-1)^j=(-1)^{L_x(L_x-1)/2}$.
Thus,
$\mathcal A_x^2 \ket 0 = \mathcal A_x \ket{\Phi} = \ket{0}$, which establishes 
\begin{equation}
\boxed{
    \mathcal A_x^2=1  \text{ for all $L_x$}.}
    \label{eqn:D4_1}
\end{equation}

The action of $\mathcal S$ on fermionic operators demonstrates 
$\mathcal S^2= e^{i\beta} T_y$ for $L_x$ even and
$\mathcal S^2=e^{i\beta'}\mathcal P T_y$ for $L_x$ odd.
We may fix these phases by considering how the vacuum state $\ket 0$ is transformed by $\mathcal S$ and $\mathcal S^2$.
The same argument used for $\mathcal A_x$ shows
$\mathcal S \ket{0} = e^{i\theta_S}\ket{\Phi}$
for the same state $\ket{\Phi}$ defined in Eq.~\eqref{eqn:phi}. Again, we set $\theta_S=0$.
Applying $\mathcal S$ to $\ket{\Phi}$ and following the same procedure used with $\mathcal A_x$ brings us to
\begin{align*}
\mathcal S \ket{\Phi} 
=&\prod_{j=0}^{L_x-1}\left[-i(-1)^j a_{L_x-1-j}\right]
\prod_{k=0}^{L_x-1} a_k^\dagger \ket 0\\
=& \begin{cases}
\ket 0 & L_x \text{ even},\\
-i \ket 0 & L_x \text{ odd}.
\end{cases}
\end{align*}
Since $\mathcal T_y$ is unitary, we can freely choose the convention $\mathcal T_y \ket 0 = \ket 0$, leading to
$\mathcal S^2 =\mathcal T_y $ for even $L_x$
and $\mathcal S^2 =-i \mathcal P \mathcal T_y$ otherwise.
Since our conventions fix $\mathcal T_y^2=\mathcal P^2 = 1$, this establishes
\begin{equation}
\boxed{
\mathcal S^4 = \begin{cases} 1 &  L_x \text{ even},\\
-1 & L_x \text{ odd}.
\end{cases}
}\label{eqn:D4_2}
\end{equation}

Finally, it can be seen on the level of fermionic operators that
$\mathcal A_x \mathcal S \mathcal A_x^{-1} =e^{i\phi} \mathcal P \mathcal S^{-1}$
where again $e^{i\phi}=\pm 1$ remains to be fixed.
The preceding calculations show that
$\mathcal A_x \mathcal S \mathcal A_x^{-1} \ket{0}=\ket \Phi$ for $L_x$ even.
Again taking $L_x$ even, applying $\mathcal S^{-1}$ to both sides of $\mathcal S\ket \Phi = \ket 0$ shows $\mathcal S^{-1}\ket 0 = \ket \Phi$, and so
$\mathcal P \mathcal{S}^{-1}\ket 0 =(-1)^{L_x}\ket{\Phi}=\ket \Phi$. In the case of odd $L_x$, we see $\mathcal A_x \mathcal S \mathcal A_x^{-1} \ket{0}=i\ket \Phi$ and $\mathcal P \mathcal{S}^{-1}\ket 0 =i\ket \Phi$. For these equalities to hold, we must have
\begin{equation}
\boxed{
\mathcal A_x \mathcal S \mathcal A_x^{-1} = \mathcal P \mathcal S^{-1} \text{ for all $L_x$}.
}\label{eqn:D4_3}
\end{equation}

The even $L_x$ cases of Eqs.~\eqref{eqn:D4_1}, \eqref{eqn:D4_2}, and \eqref{eqn:D4_3} are identical to the defining algebra of the dihedral group $D_4$  with the addition of parity appearing in Eq.~\eqref{eqn:D4_3}. Since parity $\mathcal P$ commutes with the other group generators $\mathcal A_x$ and $\mathcal S$, it forms a central extension of the group. The nontriviality of this extension is established by $(\mathcal A_x \mathcal S)^2 = \mathcal P$, where the corresponding relation in $D_4$ produces the identity.

\subsection{Sublattice parity symmetries of $H_I$}
It is simple to show any two strings of Majorana operators (that is to say, operators of the form $\gamma_1\gamma_2\dots \gamma_n$) commute so long as they share an even number of Majorana operators.
Total parity $\mathcal P$, which contains all $2N$ Majorana sites, is an example of this. This means that not only does $\mathcal P$ commute with $H$, it commutes with each term in $H$.

The plaquette interaction terms of the TLMH, $H_I$, has a number of additional symmetries generated by fermion strings not shared by $H_0$. For instance,
\begin{equation}
\mathcal P_r = \prod_{j,k} \gamma_{j,k}^r 
\quad \text{and} \quad
\mathcal P_b = \prod_{j,k} \gamma_{j,k}^b
\end{equation}
are both conserved. 
Note these are the product of Majoranas along alternating (horizontal) rows of the triangular lattice and correspond to the fermion parity of the respective sublattices thus formed. 
If we instead form alternating chains of Majoranas along the $\hat c$ direction, we arrive at another pair of operators. On our 4-leg, $y$-PBC ladder with one-dimensional index $m(i,j)=2i+j$, these are
\begin{align}
\mathcal P_\alpha =&
\prod_{m=0}^{N/4} \gr{4m}\gb{4m}\gr{4m+3}\gb{4m+3},\\
\mathcal P_\beta =&
\prod_{m=0}^{N/4}\gr{4m+1}\gb{4m+1} \gr{4m+2}\gb{4m+2}
\end{align}
where some additional care may be needed at the end of the chain on finite systems.
Finally, we can construct two more operators along the $\hat b$ direction. On our ladder, these are
\begin{subequations}
\begin{align}
\mathcal P_\mu 
=&\prod_{m=0}^{N/4}\gr{4m}\gb{4m+1}\gb{4m+2}\gr{4m+3}\\
\mathcal P_\nu 
=&\prod_{m=0}^{N/4}\gb{4m}\gr{4m+1}\gr{4m+2}\gb{4m+3},
\end{align}
\end{subequations}
again ignoring some subtleties at the ends.

It is clear these operators are related. Ignoring phases, we can see that
$\mathcal P \mathcal P_r = \mathcal P_b$,
$\mathcal P \mathcal P_\alpha = \mathcal P_\beta$, and
$\mathcal P \mathcal P_\mu = \mathcal P_\nu$. 
It can also be seen that $\mathcal P_\alpha \mathcal P_r = \mathcal P_\nu$ and $\mathcal P_\mu \mathcal P_r = \mathcal P_\beta$, which closes the algebra. Clearly, we can generate all of these operators (as well as the identity, which each of these squares to) from a set of three, for instance $\{\mathcal P, \mathcal P_r, \mathcal P_\alpha\}$.
Under the Jordan-Wigner transformation, $\mathcal P_r=i^{N/2}\prod \sy{2m-1}\sx{2m}$ and $\mathcal P_\alpha=(-1)^{N/4}\prod \sz{4m-3}\sz{4m}$.

\section{Strong-coupling limits}
\begin{figure}
    \centering
    \includegraphics[width=0.95\linewidth]{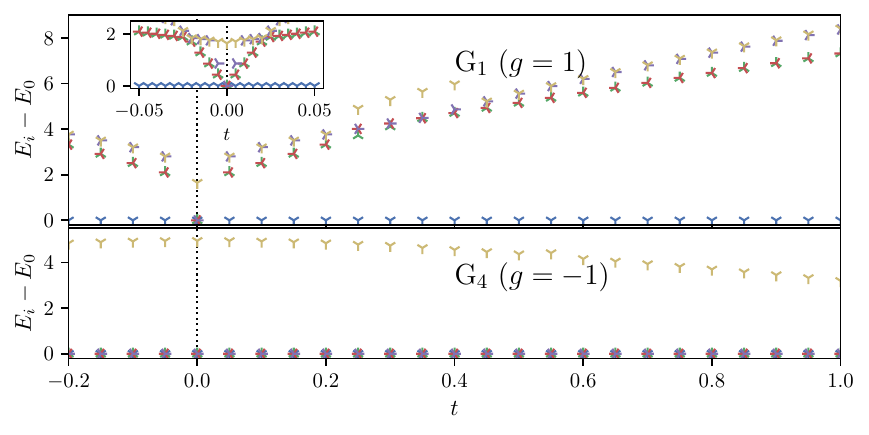}
    \caption{
    Energy gaps as a function of $t$ with fixed $g$ showing the fate of the G$_{1}$ (top) and G$_4$ (bottom) phases in the strong coupling limit. The inset zooms into the region around $t=0$ in the G$_1$ case.
    }
    \label{fig:strong}
\end{figure}

The strong-coupling ($g\rightarrow \pm \infty$) limits of the TLMH may be easily approached by fixing $g$ to a finite value and scaling $t\rightarrow 0$. To ascertain the fate of the G$_1$ phase in this limit, we begin with $g=+1$. As shown in Fig.~\ref{fig:strong}, the gap protecting this phase closes at precisely $t=0$ (equivalently, at the limit $g\rightarrow \infty$). The inset subplot shows a clear first-order transition at this point. Contrasting this, the G$_4$ phase (found by setting $g=-3$) is stable in the $t\rightarrow 0$ limit, with no gap closing or change in the fourfold ground state degeneracy anywhere in this limit.
Noting that a gauge transformation allows us to map the interaction Hamiltonian $H_I$ to $-H_I$, we should find the same phase in both limits \cite{Tummuru2021}. This is consistent with the fourfold ground state degeneracy found at $t=0$ in the $g=1$ and $g=-3$ cases, with the degeneracy occurring only at this precise point in the $g=1$ case due to a level crossing involving four states. Thus, our phase diagram in the main text indicates the G$_4$ phase occurs at both $g=\pm \infty$ limits. By examining the vicinity of this point in close detail, it is clear that the three states joining the ground state in the $g=1$ G$_1$ case correspond to high energy (rather than low-lying) excited states at $t=1$.

\section{Symmetry action on the G$_4$ ground state manifold}

To confirm that the fourfold ground state degeneracy of G$_4$ within each parity sector is associated with spontaneous breaking of $\mathcal S$ as defined in Eq.~\eqref{eqn:S}, we diagonalized the full TLMH Hamiltonian on an $L_x=8$ cylinder at $g=-3$. Using the four quasidegenerate lowest-energy eigenstates in the even parity sector, we constructed the matrix representations of the local Majorana bilinears $B_{m}=i{\gamma_{m}^b\gamma_{m+1}^r}$ within this subspace. Approximately simultaneous diagonalization of these projected operators yielded four orthonormal states $\ket{\phi_\alpha}$ with distinct staggered bulk ordering patterns.

To determine how these configurations transform under the exact antiunitary symmetry $\mathcal S$, we evaluated the matrix of squared overlaps
$Q_{\alpha\beta}
=|\langle\phi_\alpha|\mathcal S|\phi_\beta\rangle|^2$.
In the basis of the ordered states $\ket{\phi_\alpha}$, we obtain
\begin{equation}
Q\approx
\begin{pmatrix}
0.0 & 0.0001 & 0.0 & 0.9999 \\
1.0 & 0.0 & 0.0 & 0.0 \\
0.0 & 0.9999 & 0.0 & 0.0001 \\
0.0 & 0.0 & 0.9999 & 0.0
\end{pmatrix}.
\end{equation}
The dominant overlaps show $\mathcal S$ permutes the states as $\ket{\phi_0} \rightarrow \ket{\phi_1} \rightarrow \ket{\phi_2} \rightarrow \ket{\phi_3} \rightarrow \ket{\phi_0}$. The small deviations from an ideal permutation are consistent with finite-size splitting of the ground state manifold and the resulting approximate construction of the symmetry-broken states.

\section{Further details on phase transitions}

\begin{figure}
    \centering
    \includegraphics[width=0.95\linewidth]{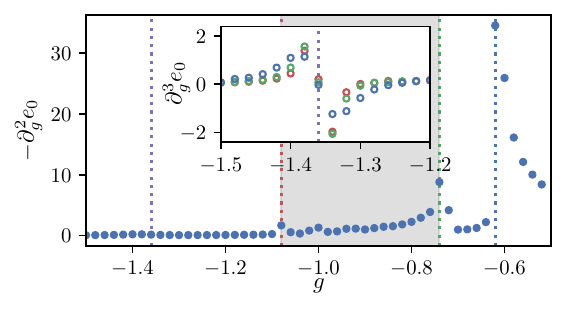}
    \caption{Energy susceptibility $-\partial_g e_0^2$ computed on an $L_x=48$ cylinder showing peaks corresponding to gap closures seen in Fig.~\ref{fig:obc_results}. Inset: third derivative of energy density showing a discontinuity at $g\approx -1.36$. Blue, green, and red points correspond to $L_x=48$, $96$, and $128$.}
    \label{fig:obc_derivs}
\end{figure}

Fig.~\ref{fig:obc_derivs} plots the second derivative of energy density computed on an $L_x=48$ cylinder. Peaks in the second derivative at $g_1\approx -0.62$, $g_2\approx -0.74$, and $g_3\approx -1.08$ occur at the same couplings as the many-body gap closures shown in Fig.~\ref{fig:obc_results}. At $g_4\approx -1.36$, no feature is apparent in the second derivative, but an apparent discontinuity at this coupling in the third derivative is shown in the inset plot.

\begin{figure}
    \centering
    \includegraphics[width=0.95\linewidth]{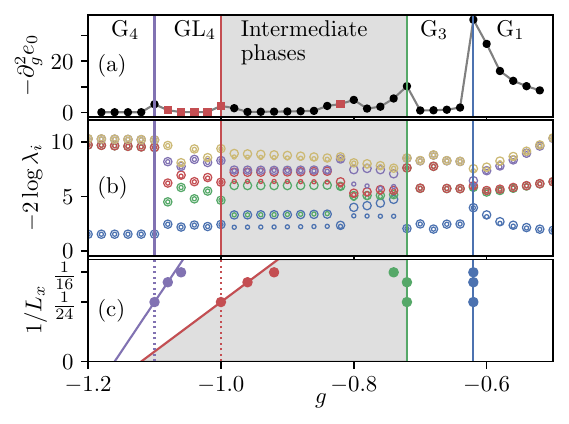}
    \caption{
    Simulation results from $x$-APBC systems.
    (a) ground state energy susceptibility from an $L_x=24$ torus. Black (red) markers correspond to regions with an even (odd) parity ground state manifold.
    (b) ground state entanglement on the same system.
    (c) Finite-size scaling of phase transition points (indicated by peaks in $-\partial_g^2 e_0$) from $x$-APBC systems with $L_x=16$, 18, and 24. The linear extrapolation is intended as a rough estimate only due to the limited range of system sizes.
    }
    \label{fig:apbc_boundary}
\end{figure}

Fig.~\ref{fig:apbc_boundary} plots the second derivative of energy (a) and entanglement spectrum (b) computed from the ground state on an $L_x=24$ $x$-APBC torus. Subplot (c) shows the location of the observed phase transitions as a function of $L_x$, with a linear extrapolation presented as a rough estimate of the phase boundaries in the thermodynamic limit. As these results were obtained for only $L_x=16$, 18, and 24, this extrapolation is very approximate, leading to inconsistencies with the $x$-OBC results (in which system sizes of up to $L_x=128$ were considered).

\section{G$_3$ and GL$_4$ phases}
\begin{figure}
    \centering
    \includegraphics[width=0.95\linewidth]{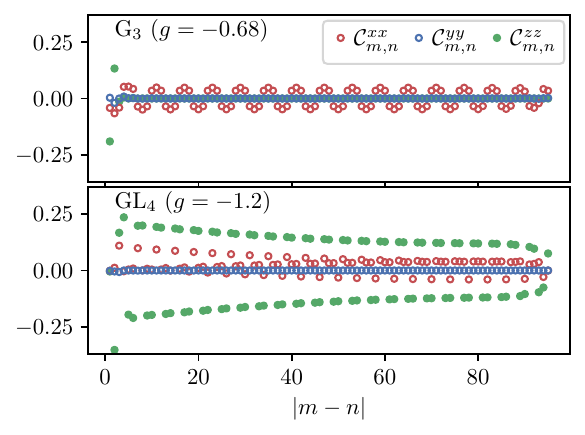}
    \caption{Magnetic correlations in the G$_3$ and GL$_4$ phases computed on an $L_x=96$ cylinder in the even parity sector. One of four degenerate states was selected in the GL$_4$ phase.}
    \label{fig:g3g4}
\end{figure}

Fig.~\ref{fig:g3g4} plots magnetic and fermionic correlations in the G$_3$ and GL$_4$ phases computed on a cylinder with $L_x=96$. This plot makes clear a six-site repeating structure in the G$_3$ phase. Correlations in the GL$_4$ phase are less easily understood. Notably, this phase exhibits long-range connected correlations in $\sigma^z$, which are absent in the gapped SPT phases.

\begin{figure}
\includegraphics[width=.95\linewidth]{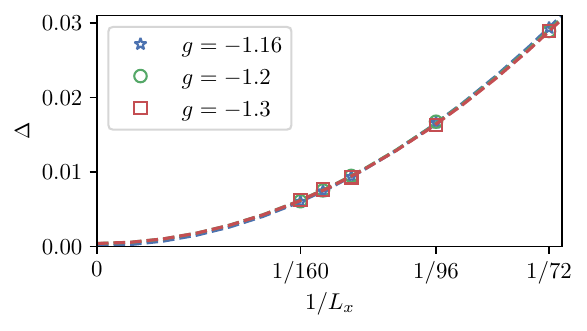}
\caption{Finite-size scaling of gaps above the ground state manifold in the GL$_4$ phase computed on cylinders with even fermion parity. The plotted best-fit lines are of the form $\Delta = \alpha/L_x^2 + \beta$ with $\alpha\approx 150$ and very small $\beta\sim 10^{-4}$.}
\label{fig:gl4gap}
\end{figure}

Like the G$_4$ phase discussed in the main text, the GL$_4$ phase exhibits a fourfold ground state degeneracy on tori that doubles to eightfold on cylinders. However, unlike the G$_4$ phase, in which the gap $\Delta$ separating the ground state manifold from the remainder of the spectrum remains finite as $L_x\rightarrow \infty$, the GL$_4$ phase exhibits a gapless spectrum. Finite-size scaling on $x$-OBC systems, shown in Fig.~\ref{fig:gl4gap}, indicates $\Delta\propto 1/L_x^2$ as $L_x\rightarrow \infty$.

\section{Additional results from odd-$L_x$ tori}

\begin{figure}
\includegraphics[width=.95\linewidth]{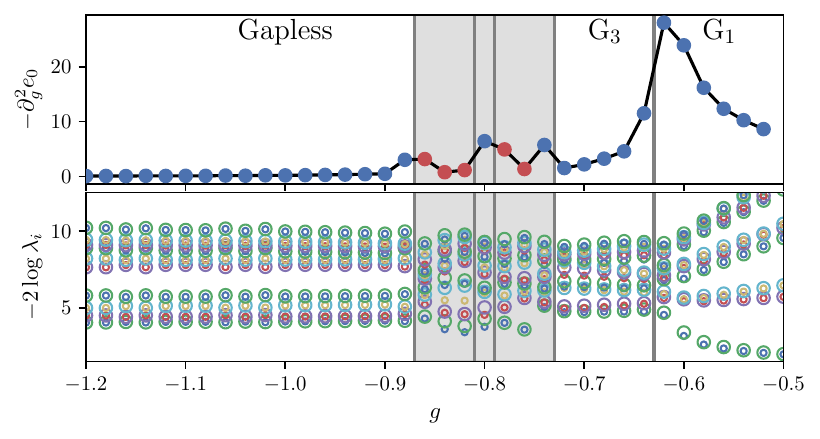}
\caption{Energy susceptibility (top) and entanglement spectrum as a function of $g$ on an $x$-APBC torus with $L_x=19$. Vertical lines indicate phase transitions. Blue (red) circles on the upper plot indicate even (odd) parity ground states.}
\label{fig:19_phase}
\end{figure}

By considering only the odd-$L_x$, $x$-APBC case in DMRG simulations, previous work on the 4-leg TLMH found a gapless phase  \cite{Tummuru2021} in place of the gapped G$_4$ phase discussed in the main body of this text.
Our DMRG simulations on odd-$L_x$ systems suggest that the gapless mode occurs on tori regardless of $x$ boundary condition (PBC or APBC), while odd-$L_x$ cylinders instead host the gapped phases seen elsewhere. 
Here, we share some of our results on odd-$L_x$, APBC systems.

Fig.~\ref{fig:19_phase} plots energy susceptibility and entanglement spectrum computed from an $L_x=19$ $x$-APBC torus. As indicated in the figure, we observe the G$_1$ and G$_3$ phases as before. This is followed by a difficult-to-analyze region with a number of apparent phase transitions. Following this, we observe a single phase for the range of couplings $g$ that are split between GL$_4$ and G$_{4}$ phases in other geometries. While prior work describes a unique ground state in this region for odd $L_x$ \cite{Tummuru2021}, our results indicate a twofold degeneracy.

\begin{figure}
\includegraphics[width=.95\linewidth]{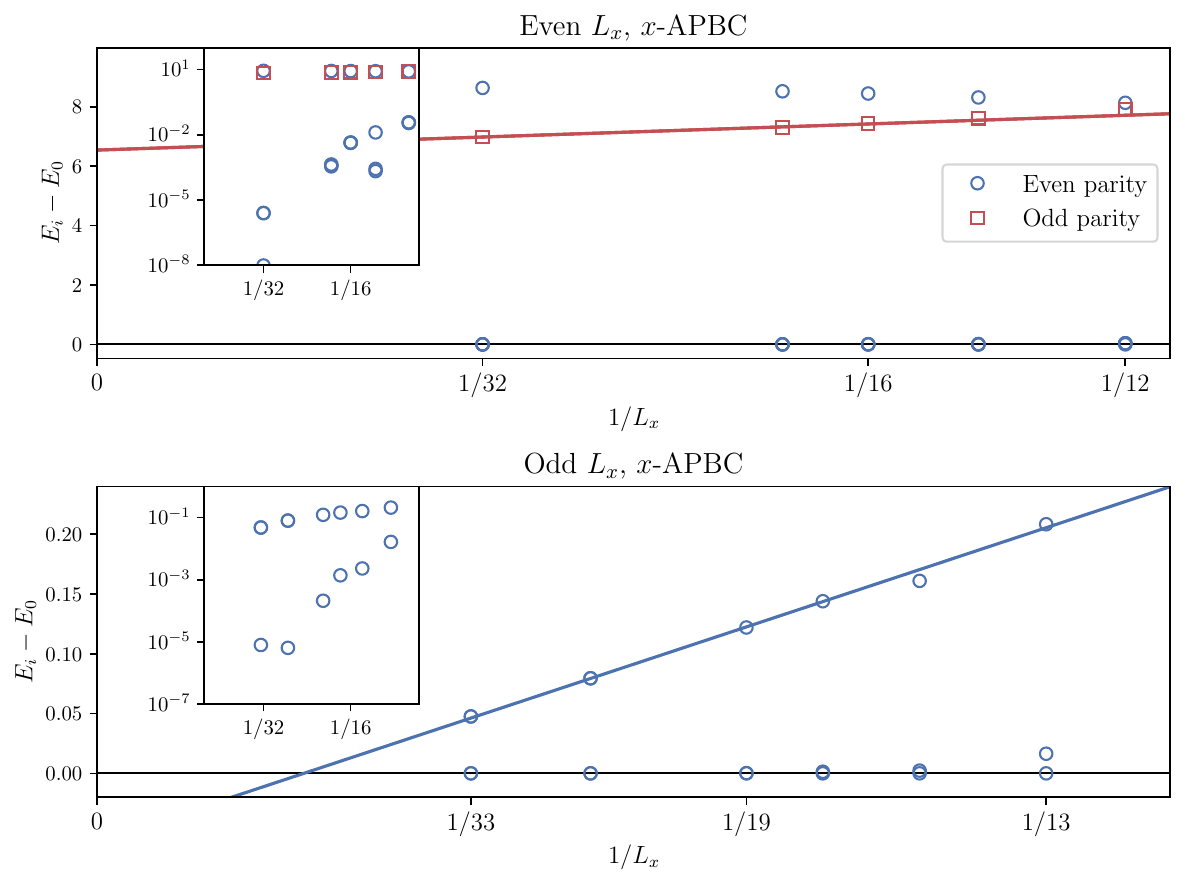}
\caption{Finite-size scaling of low-lying gaps at $g=-3$ on tori with $x$-APBC for $L_x$ even (top) and odd (bottom). Note the different $y$-axis scales. Red squares on the top figure correspond to odd-parity states, while circles indicate even-parity states. Insets plot the same data with gaps on a logarithmic scale to show the four and twofold quasi-degeneracy of the ground state for even and odd $L_x$, respectively.
}
\label{fig:gap_scaling}
\end{figure}

Analysis of gaps in this phase shows $1/L_x$ scaling consistent with a relativistic gapless mode. Fig.~\ref{fig:gap_scaling} compares these gaps (computed at $g=-3$) with those in even-length tori (which are in the G$_4$ phase for this value of $g$). Both our results and those presented in earlier work \cite{Tummuru2021} have a strange feature: the $1/L_x$ gap scaling extrapolates to a negative, finite gap as $L_x\rightarrow \infty$ rather than zero gap as should be the case. While this would seem most likely to be a finite-size effect of some sort, difficulties in DMRG simulations (resulting from both the required periodic boundaries, the highly entangled nature of the eigenstates, and the twofold degeneracy of the ground state) prevent us from obtaining well-converged excited states required to compute these gaps for $L_x>33$. 

\begin{figure}
\includegraphics[width=.95\linewidth]{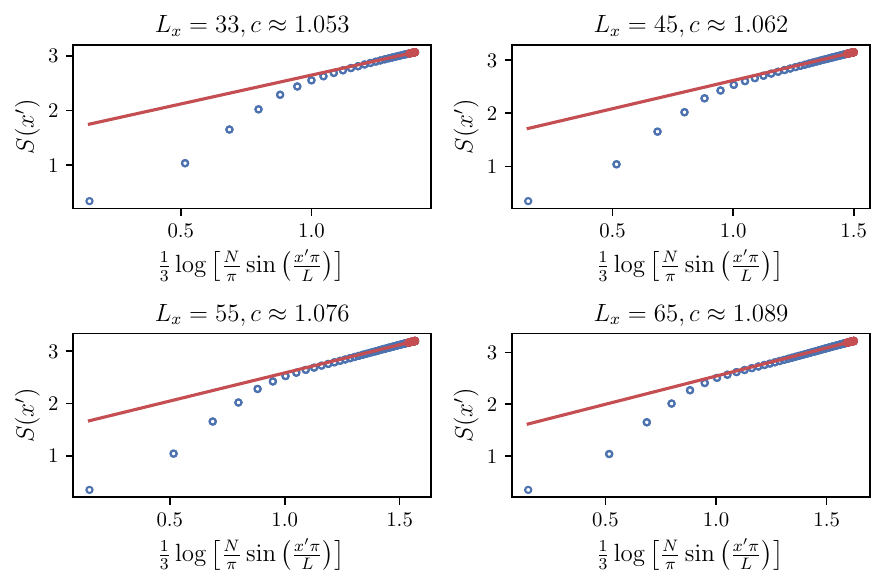}
\caption{Entanglement entropies for odd-$L_x$, APBC systems at $g=-3$ showing $c\approx 1$. Points used in obtaining the fits are shown as red circles.
}
\label{fig:big_charges}
\end{figure}

Further support of the gapless nature of this phase can be found by examining its central charge $c$. We follow the approach taken in earlier work \cite{Tummuru2021}, which assumes the phase can be described by a 1+1-dimensional CFT on our $N$-site lattice. In this case, the entanglement entropy for a bipartition at the $x$th site is predicted to fit the formula
\begin{equation}
S(x)=\frac{c}{3}\log\left[\frac{N}{\pi}\sin\left(\frac{x\pi}{N}\right)\right] + S_0
\label{eqn:s_charge}
\end{equation}
where $S_0$ is a non-universal constant \cite{Calabrese2009}. To account for an even-odd effect, we replace $S(x)$ with the entanglement entropy averaged over two adjacent sites, $S(x')=\frac{1}{2}\left(S(x)+S(x+1)\right)$ with $x'=x+\frac{1}{2}$ \cite{Tummuru2021}. In practice, we obtain $c$ by obtaining a least squares fit of $S(x')$ as a function of
$\ell(x')=\frac{1}{3}\log\left[\frac{N}{\pi}\sin\left(\frac{x\pi}{N}\right)\right]$, with the data restricted to the largest few values of $\ell$ to avoid short-range physics. 

Entanglement entropies and best-fit lines indicating central charges of $c\approx 1$ at $g=-3$ (in the gapless phase) are shown in systems ranging from $L_x=33$ to $L_x=65$. In all cases, bond dimensions were chosen to maintain a DMRG truncation error of at most $10^{-6}$, with the $L_x=65$ case requiring a bond dimension of $m=2500$. In all cases, values of $c\approx 1$ were obtained.

\begin{figure}
\includegraphics[width=.95\linewidth]{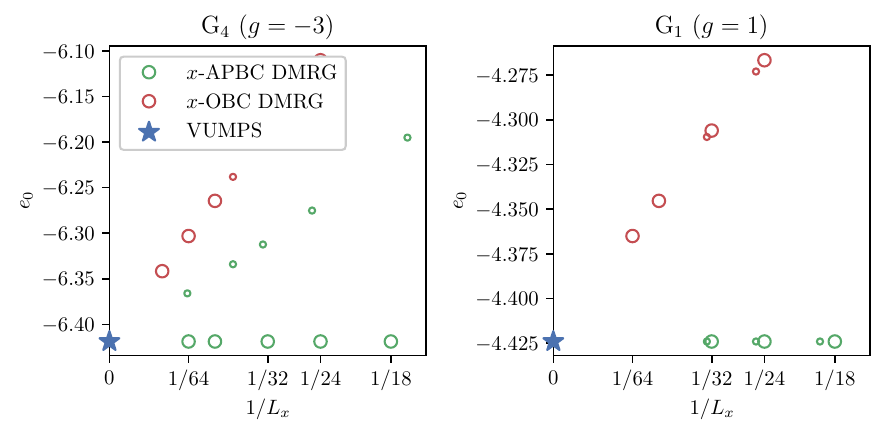}
\caption{Comparison of energy density between $x$-APBC and $x$-OBC DMRG results for even and odd-$L_x$ systems (denoted with large and small circles, respectively). The star at $1/L_x=0$ corresponds to the infinite-system energy density obtained by VUMPS.}
\label{fig:energy_comparison}
\end{figure}

As discussed in the main text, we hypothesize the gapless phase is an anomaly most likely caused by a frustration-induced, deconfined domain wall. Comparison of energy densities $e_0 =\frac{1}{2 L_x}E_0$ in Fig.~\ref{fig:energy_comparison} from different boundary conditions and system sizes indicates this gapless mode corresponds to a higher-energy ground state than similar-sized even-$L_x$ periodic systems, with energy difference shrinking as $1/L_x$ in the thermodynamic limit and in all cases extrapolating to the value of energy density obtained from VUMPS for an infinite-MPS ground state. 

This scaling in energy density is consistent with the formation of a domain wall (or other defect) at a cost of $\mathcal O(1)$ in energy, thus leading to a $1/L_x$ penalty in the energy density.
A similar energy penalty occurs for $x$-OBC systems with $L_x$ even or odd, which can be ascribed to the absence of $\mathcal O(1)$ terms in the Hamiltonian connecting the ends of the cylinder. The lack of these energy-reducing terms causes a $1/L_x$ increase in energy density.

\section{Mean-field theory}
\label{app:MFT}
Earlier work on the TLMH developed a mean-field theory well-suited for the full two-dimensional lattice \cite{Tummuru2021}. As a simplification, they considered a state in which fermion bilinears appearing in the Hamiltonian, $\braket{i \gamma_i \gamma_j}$,
can assume one of four values: $\Delta_2$ for all next-nearest neighbors, $\Delta_1$ for nearest-neighboring sites in the $a$ and $b$ directions (as illustrated in Fig.~\ref{fig:bilinears}), and $\Delta_c$ or $\Delta_{\overline{c}}$ for alternating sites in the $c$ direction. This choice allows for nearest-neighbor dimerization along a single favored axis.
Using this mean-field, the authors obtained a phase diagram consisting of two gapped topological insulators with Chern numbers $\mathcal C=1$ for $g > g_c^{\text{MF}}\approx-0.73$ and $\mathcal C=3$ for $g < g_c^\text{MF}$. 

\begin{figure}
\centering
\begin{tikzpicture}
\begin{scope}[thick,decoration={
    markings,
    mark=at position 0.5 with {\arrow{stealth}}}
    ] 
    \draw [postaction=decorate] (0,0) -- (1,0);
    \draw [postaction=decorate] (-0.5, {0.5*sqrt(3)}) -- (0,0);
    \draw [postaction=decorate] (0,0) -- (-0.5, {-0.5*sqrt(3)});
    \draw [postaction=decorate] (0,0) -- (0, {sqrt(3)});
    \draw [postaction=decorate] (0,0) -- (-1.5, {-0.5*sqrt(3)});
    \draw [postaction=decorate] (0,0) -- (1.5, {-0.5*sqrt(3)});
\end{scope}
    \filldraw[red] (0,0)  circle (2pt);
    \node at (.5,.2) {$\Delta_{ a}$};
    \filldraw[red] (1,0)  circle (2pt);
    \filldraw[blue] (-0.5, {0.5*sqrt(3)})  circle (2pt);
    \filldraw[blue] (-0.5, {-0.5*sqrt(3)})  circle (2pt);
    \filldraw[red] (0, {sqrt(3)}) circle (2pt);
    \node at (0.25, 1) {$\Delta_{\alpha}$};
    \filldraw[blue] (-1.5, {-0.5*sqrt(3)})  circle (2pt);
    \filldraw[blue] (1.5, {-0.5*sqrt(3)})  circle (2pt);
\end{tikzpicture}
\begin{tikzpicture}
\begin{scope}[thick,decoration={
    markings,
    mark=at position 0.5 with {\arrow{stealth}}}
    ] 
    \draw [postaction=decorate] (1,0) -- (0,0);
    \draw [postaction=decorate] (-0.5, {0.5*sqrt(3)}) -- (0,0);
    \draw [postaction=decorate] (-0.5, {-0.5*sqrt(3)}) -- (0,0);
    \draw [postaction=decorate] (0, {sqrt(3)}) -- (0,0);
    \draw [postaction=decorate] (0,0) -- (-1.5, {-0.5*sqrt(3)}) ;
    \draw [postaction=decorate] (1.5, {-0.5*sqrt(3)}) -- (0,0);
\end{scope}
    \filldraw[blue] (0,0)  circle (2pt);
    \node at (.5,.2) {$\Delta_{\overline a}$};
    \filldraw[blue] (1,0)  circle (2pt);
    \filldraw[red] (-0.5, {0.5*sqrt(3)})  circle (2pt);
    \filldraw[red] (-0.5, {-0.5*sqrt(3)})  circle (2pt);
    \filldraw[blue] (0, {sqrt(3)}) circle (2pt);
    \node at (0.25, 1) {$\Delta_{\overline \alpha}$};
    \filldraw[red] (-1.5, {-0.5*sqrt(3)})  circle (2pt);
    \filldraw[red] (1.5, {-0.5*sqrt(3)})  circle (2pt);
\end{tikzpicture}
\caption{Naming conventions for the fermion bilinears used to form the mean-field theory. 
The remaining directions ($b$, $c$ for nearest and $\beta$, $\gamma$ for next-nearest) are labeled counter-clockwise.
}
\label{fig:bilinears}
\end{figure}

On the ladder geometries we consider, the choice of dimerization axis is no longer arbitrary. We also allow for different (and alternating) values of next-nearest neighbor bilinears, leading to a set of 12
mean-field parameters.
A mean-field (Slater determinant) wavefunction $\ket{\text{MF}}$ of this form has energy density
\begin{align}
\frac{\braket{H}}{N}=& t\l(\Delta_a + \Delta_{\overline{a}} + \Delta_b + \Delta_{\overline{b}} + \Delta_c + \Delta_{\overline{c}}\r) \nonumber\\
&-g\l[4 \Delta_a\Delta_{\overline{a}} + \l(\Delta_b + \Delta_{\overline{b}}\r)^2 + \l(\Delta_c + \Delta_{\overline{c}}\r)^2\r]\nonumber\\
&+g\l(\Delta_a\Delta_{\overline{\alpha}} + \Delta_{\overline{a}}\Delta_\alpha 
      + \Delta_b\Delta_{\overline{\beta}} \r)\nonumber\\
&+g\l( \Delta_{\overline{b}}\Delta_\beta + \Delta_c\Delta_\gamma + \Delta_{\overline{c}}\Delta_{\overline{\gamma}}\r).
      \label{eqn:sd_expt}
\end{align}
A correctly optimized state $\ket{\text{MF}}$ satisfies
$\frac{1}{N}\frac{\partial}{\partial \tau_j} \braket{H}=0$
for mean-field parameters $\tau_j$ that will be introduced shortly and with $j\in \{a,b,c,\ola,\olb,\olc,\alpha,\beta,\gamma,\overline{\alpha},\overline{\beta},\overline{\gamma}\}$.
These derivatives can be written explicitly:
\begin{align}
\frac{1}{N}\frac{\partial \braket{H}}{\partial \tau_j}=&
\l(t-4g\Delta_{\overline{a}}+g\Delta_{\overline{\alpha}}\r)\frac{\partial\Delta_a}{\partial\tau_j} \nonumber \\
& + \l(t-4g\Delta_a+g\Delta_\alpha\r)\frac{\partial\Delta_{\overline{a}}}{\partial\tau_j} \nonumber\\
&+\l[t-2g\l(\Delta_b+\Delta_\olb\r)+g\Delta_{\overline{\beta}}\r]\frac{\partial\Delta_b}{\partial\tau_j} \nonumber \\
&+\l[t-2g\l(\Delta_b+\Delta_\olb\r)+g\Delta_\beta\r]\frac{\partial\Delta_\olb}{\partial\tau_j}\nonumber\\
&+\l[t-2g\l(\Delta_c+\Delta_\olc\r)+g\Delta_\gamma\r]\frac{\partial\Delta_c}{\partial\tau_j} \nonumber \\
&+\l[t-2g\l(\Delta_c+\Delta_\olc\r)+g\Delta_{\overline{\gamma}}\r]\frac{\partial\Delta_\olc}{\partial\tau_j}\nonumber\\
&+g\Delta_\ola\frac{\partial\Delta_\alpha}{\partial\tau_j}
 +g\Delta_a\frac{\partial\Delta_{\overline{\alpha}}}{\partial\tau_j}\nonumber \\
&+g\Delta_\olb\frac{\partial\Delta_\beta}{\partial\tau_j} 
 +g\Delta_b\frac{\partial\Delta_{\overline{\beta}}}{\partial\tau_j}\nonumber \\
&+g\Delta_c\frac{\partial\Delta_\gamma}{\partial\tau_j}
 +g\Delta_\olc\frac{\partial\Delta_{\overline{\gamma}}}{\partial\tau_j}\label{eqn:e_deriv}.
\end{align}
To fix the parameters $\tau_j$, we form a quadratic mean-field Hamiltonian $H_\text{MF}$ such that the ground state of $H_\text{MF}$ is the optimal Slater determinant minimizing Eq.~\eqref{eqn:sd_expt}.
This mean-field Hamiltonian is written
\begin{align}
    H_{\text{MF}}
= i\sum_{m,n}\bigg[&\tau_a\gamma_{m,n}^r\gamma_{m+1,n}^r
                    +\tau_\ola \gamma_{m+1,n}^b\gamma_{m,n}^b \\
                    &+\tau_b\gamma_{m-1,n}^b\gamma_{m,n}^r
                    +\tau_\olb \gamma_{m,n+1}^r\gamma_{m,n}^b \nonumber\\
                    &
                    +\tau_c \gamma_{m,n}^r\gamma_{m-1,n-1}^b
                    +\tau_\olc \gamma_{m,n}^r\gamma_{m,n}^b \nonumber \\
        &+\tau_\alpha\gamma_{m,n}^r\gamma_{m,n+1}^r 
        + \tau_{\overline{\alpha}}\gamma_{m,n+1}^b\gamma_{m,n}^b \nonumber \\
        &+\tau_\beta\gamma_{m,n}^r\gamma_{m-2,n-1}^b 
        +\tau_{\overline{\beta}}\gamma_{m,n}^b\gamma_{m-1,n}^r \nonumber \\
        &+ \tau_\gamma\gamma_{m,n}^r\gamma_{m+1,n-1}^b 
        + \tau_{\overline{\gamma}}\gamma_{m+2,n}^r\gamma_{m,n}^b\bigg].\nonumber 
\end{align}
with expectation value $\frac{1}{N}\braket{H_\text{MF}}=e_{\text{MF}}=\sum_j \tau_j \Delta_j$.
To form self-consistency equations, we take derivatives of $e_{\text{MF}}$ with respect to the mean-field couplings:
\begin{equation}
 \frac{\partial e_\text{MF}}{\partial \tau_j} = \Delta_a + \sum_{k\neq j}\tau_k\frac{\partial \Delta_k}{\tau_k}. \label{eqn:sc1}
\end{equation}
The Hellmann-Feynmann theorem allows us to identify these derivatives with the bilinear expectation values $\Delta_j$:
\begin{equation}
\Delta_j = \frac{1}{N}\frac{\partial}{\partial \tau_j}\braket{H_\text{MF}}
\equiv  \frac{\partial e_\text{MF}}{\partial \tau_j}.
\label{eqn:sc2}
\end{equation}
For Eqs. \eqref{eqn:sc1} and \eqref{eqn:sc2} to agree, we must have
\begin{equation}
    \sum_{k\neq j}\tau_k \frac{\partial \Delta_k}{\tau_k}=0,
    \label{eqn:sc3}
\end{equation}
again with $j\in \{a,b,c,\ola,\olb,\olc,\alpha,\beta,\gamma,\overline{\alpha},\overline{\beta},\overline{\gamma}\}$.
Matching the coefficients of each derivative $\partial \Delta_j/\partial \tau_j$ between \eqref{eqn:sc3} and \eqref{eqn:e_deriv} leads to the self-consistency equations
\begin{align}
\tau_a=&t-4g\Delta_\ola+g\Delta_{\overline{\alpha}}, \nonumber \\
 \tau_\ola =& t-4g\Delta_a + g \Delta_\alpha,
\nonumber\\
\tau_b =& t-2g(\Delta_b+\Delta_\olb) + g \Delta_{\overline{\beta}}  \nonumber \\
\tau_\olb =& t-2g(\Delta_b+\Delta_\olb) + g \Delta_{\beta}, 
\nonumber\\
\tau_c =& t-2g(\Delta_c+\Delta_\olc) + g \Delta_\gamma, \nonumber \\
\tau_\olc =&t-2g(\Delta_c+\Delta_\olc) + g \Delta_{\overline{\gamma}}, 
\nonumber\\
\tau_\alpha =& g\Delta_\ola, \qquad \tau_{\overline{\alpha}} = g\Delta_a ,
\nonumber\\
\tau_\beta =& g\Delta_\olb, \qquad \tau_{\overline{\beta}} = g\Delta_b,
\nonumber\\
\tau_\gamma =& g\Delta_c, \qquad \tau_{\overline{\gamma}} = g\Delta_\olc.
\label{eqn:sc_full}
\end{align}

The final ingredient required to solve this set of self-consistency equations is a momentum-space representation of the single-particle Hamiltonian.
In momentum space, Majoranas can be shown to obey $\gamma_{-\mathbf k}=\gamma_{\mathbf{k}}^\dagger$. Thus, momenta $\mathbf{k}$ and $\mathbf{-k}$ are not distinct and so sums and integrals over momenta should be taken over one-half of the Brillouin zone, indicated by notation $\sum_\textbf{k}'$ \cite{Tummuru2021}. With this taken into account, we make use of notation $\Psi_\mathbf{k}=\begin{pmatrix}\psi_\mathbf{k}^r \\ \psi_\mathbf{k}^b\end{pmatrix}$
to write the mean-field Hamiltonian
\begin{align}
H_\text{MF} =& \sum_\mathbf{k}' \Psi_\mathbf{k}^\dagger
\begin{pmatrix}
D_1(\mathbf{k}) & D_2(\mathbf{k}) \\
D_2^*(\mathbf{k}) & -D_{\overline{1}}(\mathbf{k})
\end{pmatrix}
\Psi_\mathbf{k}
\label{eqn:hk}
\end{align}
 and
\begin{subequations}
\begin{align}
D_1(\mathbf{k}) = -4\big(&\tau_a \sin k_x + \tau_\alpha \sin k_y\big) \\
D_{\overline{1}}(\mathbf{k}) =-4\big(&\tau_\ola \sin k_x + \tau_{\overline{\alpha}} \sin k_y\big)\\
D_2(\mathbf{k}) =2i\big(& -\tau_b e^{-ik_x} + \tau_\olb e^{-ik_y} \\
&+ \tau_c e^{-i(k_x+k_y)}+\tau_\olc \nonumber \\
&+\tau_\beta e^{-i(2k_x+k_y)} - \tau_{\overline{\beta}} e^{ik_x} \nonumber \\
&+\tau_\gamma e^{i(k_x-k_y)}+\tau_{\overline{\gamma}}e^{-2ik_x}
\big).\nonumber
\end{align}
\end{subequations}
The single-particle spectrum $\epsilon_{\text{MF}}^\pm(\mathbf k)$ is then given by
\begin{align}
\epsilon_{\text{MF}}^\pm(\mathbf k)=&\frac{D_1(\mathbf k) - D_{\overline 1}(\mathbf k)}{2} \\
& \pm \sqrt{\left[\frac{D_1(\mathbf k)+D_{\overline{1}}(\mathbf k)}{2}\right]^2 + |D_2(\mathbf k)|^2} \nonumber
\end{align}
with ground state mean-field energy density
$e_\text{MF} = -\frac{1}{2}\sum_{\mathbf k}'\left(|\epsilon^+_\text{MF}(\mathbf k)| + |\epsilon^-_\text{MF}(\mathbf k)|\right)$.

As expected, the Hamiltonian \eqref{eqn:hk} and self-consistency equations \eqref{eqn:sc_full} reduce to the simpler mean-field theory used in Ref.~\cite{Tummuru2021} when we restrict $\tau_a=\tau_\ola=\tau_b=\tau_\olb\equiv \tau_1$ and $\tau_\alpha=\tau_{\overline{\alpha}}=\tau_\beta=\dots\equiv \tau_2$. As such, the more general mean-field theory will have an energy density bounded from above by the simpler version.

\subsection{Self-consistency loop}
To obtain self-consistent values of the mean-field parameters $\tau_j$, we utilize the following algorithm:
\begin{enumerate}
\item Choose a starting set of values for $\tau_j$.
\item Use derivatives of \eqref{eqn:hk} to obtain corresponding values of $\Delta_j$ from Eq.~\eqref{eqn:sc2}.
\item Use \eqref{eqn:sc_full} to obtain new values of $\tau_j$.
\item Repeat steps 2 and 3 until quantities converge within a desired precision.
\end{enumerate}
This procedure can be adapted to infinite systems by replacing sums over $k$ with integration.

We restrict ourselves to the same 4-leg ladder used in our DMRG simulations, which is the $L_y=2$,\footnote{In our notation, $L_y$ corresponds to the number of Dirac modes (unit cells) along the $y$ direction, not the number of rows of the underlying triangular lattice.} 
$y$-PBC case of a general $L_x\times L_y$ parallelogram on the rectangular lattice with
a Brillouin zone consisting of points $\mathbf k = (k_x, k_y)$
with $k_x =  \frac{j\pi}{L_x}$ for
\begin{equation*}
 j=\begin{cases}
-L_x, -L_x+2,\dots, L_x-2 & x\text{-PBC}\\
-L_x+1, -L_x+3,\dots, L_x-1 & x\text{-APBC}\\
\end{cases}
\end{equation*}
and $k_y =  \frac{j\pi}{\sqrt 3 L_y}$ for
\begin{equation*}
j=\begin{cases}
-L_x, -L_x+2,\dots, L_y-2 & y\text{-PBC}\\
-L_x+1, -L_x+3,\dots, L_y-1 & y\text{-APBC}.
\end{cases}
\end{equation*}

\subsection{Mean-field results}

With our generalized mean-field theory, we obtain a similar phase diagram to that in \cite{Tummuru2021}, in that we observe two gapped phases with a transition in a similar region. For $g > g_c^{\text{MF}}$, we observe no dimerization and thus our results reduce exactly to the simpler original mean-field theory. Below $g_c$, the generalized mean-field obtains a slightly lower energy.

\begin{figure}[h]
\includegraphics[width=0.95\linewidth]{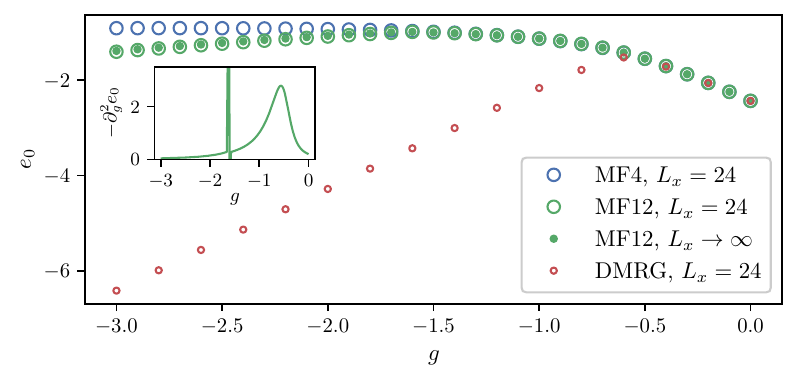}
\caption{Comparison of energy densities between the four-parameter mean-field (MF4) used in earlier work \cite{Tummuru2021}, the 12-parameter mean-field derived here (MF12) and DMRG results, all computed on finite ladders with $x$-APBC. $L_x\rightarrow \infty$ mean-field results were obtained from finite-size scaling. The inset plots energy susceptibility for the $L_x\rightarrow\infty$ MF12 case.
}
\label{fig:mf_energy}
\end{figure}

Comparison with DMRG results, with both DMRG and mean-field computed on 4-leg ladders, is shown in Fig.~\ref{fig:mf_energy}. While the energy densities coincide in the G$_1$ phase (indicating agreement between mean-field and exact results), the DMRG results obtain a much lower energy for $g < g_1^{\text{MF}}\approx -0.5$.\footnote{
Note that while finite-size scaling results for DMRG are not presented in this figure, they are shown in Fig.~\ref{fig:energy_comparison} at $g=1$ and $g=-3$, with both cases indicating a minimal change in energy density ($\mathcal O(10^{-3}$) as opposed to the $\mathcal O(1)$ discrepancy observed between finite-size exact (DMRG) energy density and either finite-size or infinite-system mean-field energy densities for $g<g_1$.}

It should be noted that this 12-parameter ladder mean-field theory undergoes a first-order phase transition at $g\approx -1.5$ (signaled by cusp in energy that causes a singularity in $-\partial_g^2 e_0$ as shown in Fig.~\ref{fig:mf_energy}). This is in addition to the continuous phase transition at $g\approx -0.5$ found in the simpler four-parameter mean-field theory \cite{Tummuru2021}.

\newpage
\bibliography{sources}

\end{document}